**Probabilistic Prediction for Binary Treatment Choice: with focus on personalized medicine**

Charles F. Manski

Department of Economics and Institute for Policy Research, Northwestern University

October 2021

Abstract

This paper extends my research applying statistical decision theory to treatment choice with sample data, using maximum regret to evaluate the performance of treatment rules. The specific new contribution is to study *as-if optimization* using estimates of illness probabilities in clinical choice between surveillance and aggressive treatment. Beyond its specifics, the paper sends a broad message. Statisticians and computer scientists have addressed conditional prediction for decision making in indirect ways, the former applying classical statistical theory and the latter measuring prediction accuracy in test samples. Neither approach is satisfactory. Statistical decision theory provides a coherent, generally applicable methodology.

I have benefitted from the opportunity to present this work at the Northwestern econometrics seminar. I am grateful for comments from Michael Gmeiner, Valentyn Litvin, and Filip Obradovic. I am grateful to Litvin and Gmeiner for programming the computations in Sections 4.1 and 4.2 respectively.

## 1. Introduction

A classic concern of probability theory and statistics has been to predict realizations of a real random variable y conditional on realizations of a covariate vector x. A standard formalization of the problem begins with a population characterized by a joint distribution P(y, x). A member is drawn at random from the sub-population with a specified value of x. The problem is to predict y conditional on x. The conditional distribution P(y|x) provides the complete feasible probabilistic prediction.

Rather than study P(y|x) in totality, researchers often focus on a real-valued feature of P(y|x) that is interpretable as a *best point predictor* of y conditional on x. The standard approach has been to minimize the conditional expected loss from prediction errors, with respect to a given loss function. Thus, one solves a minimization problem $\min_{p \in (-\infty, \infty)} E[L(y - p)|x]$, where p is a predictor value, $y - p$ is prediction error, and $L(\cdot)$ is the loss function. Familiar cases include square and absolute loss, yielding the conditional mean and median as best predictors. For expositions, see Ferguson (1967) and Manski (2007a).

A standard formalization of the statistical problem supposes that one does not know P(y, x). Instead, one observes $(y_i, x_i, i = 1, \ldots, N)$ in a random sample of N persons drawn from a study population that has distribution P(y, x). One uses the sample data to estimate P(y|x), or a best point predictor.

The standard formalization considers probabilistic prediction as a self-contained problem, without reference to an external application. Obtaining a best point prediction by minimizing expected loss solves a decision problem that may have applications. However, the commonly used loss functions, notably square and absolute loss, have usually been motivated by tradition and tractability rather than by losses incurred in actual decisions that require choice of point predictions.

This paper considers probabilistic predictions used to inform decisions in an important class of applications, namely choice between two treatments for members of a heterogeneous population. To flesh out abstract ideas, I analyze choice between surveillance and aggressive treatment in personalized medicine. Building on earlier work in Manski (2018, 2019a), I consider a clinician caring for patients with observed covariates x. The setting supposes that there are two care options for a specified disease, with A denoting surveillance and B denoting aggressive treatment. Let y = 1 if a patient is ill with the disease and y = 0 if not. I pose a model of patient welfare in which surveillance is the better option if y = 0 and aggressive treatment if y = 1.

The analysis in this paper applies as well to other decision problems that have the same mathematical structure. One such is judicial treatment of criminal defendants. Here the choice is to find a defendant guilty or not guilty of a crime. A defendant is analogous to a patient. A guilty decision is analogous to aggressive treatment, and a not-guilty one is analogous to surveillance. Uncertainty about whether a defendant committed the crime is analogous to uncertainty about whether a patient is ill. See Manski (2020) for discussion relating judicial and clinical decisions.

Decision making would be simple if y were observable at the time of treatment choice. However, the clinician must choose without knowing the illness status of the patient. This generates a rationale to predict illness. Given knowledge of x, the most that a clinician can do is to learn $P(y = 1|x)$. The model of patient welfare implies that surveillance is the better option if $P(y = 1|x)$ is less than a known threshold $p_x^*$ and aggressive treatment is better if $P(y = 1|x)$ exceeds $p_x^*$. Hence, the best point prediction is y = 1 if $P(y = 1|x) > p_x^*$ and y = 0 if $P(y = 1|x) < p_x^*$.

Empirical research on medical risk assessment has used sample data on illness in study populations to estimate conditional probabilities of illness or to make point predictions of illness. Risk assessment has long been performed by biostatisticians who use classical frequentist statistical theory to propose inferential methods and assess findings. Although the motivation may be to improve patient care, biostatistical analysis commonly views prediction as a self-contained inferential problem rather than as a task undertaken specifically to inform treatment choice.

In the 21$^{st}$ century, medical risk assessment is increasingly performed by computer scientists, who view prediction methods as computational algorithms rather than as approaches to statistical inference. Frequentist statisticians maintain an ex-ante perspective, studying how methods perform across repetitions of a sampling process. In contrast, computer scientists perform ex post evaluation, fitting an algorithm on a "training" sample and examining the accuracy of the predictions it yields on a "test" sample. Breiman (2001) argues for this approach, writing (p. 201): "Predictive accuracy on test sets is the criterion for how good the model is." Breiman does not explain why this is or should be the criterion. He just states it.

Measuring performance in this ex-post manner may have appeal to clinicians who lack expertise in statistical methodology but who may feel that they can appraise ex post prediction accuracy heuristically. However, by its nature, evaluation on a test sample cannot yield lessons that generalize beyond the particular test performed. Efron (2020), in an article contrasting the perspectives of statisticians and computer scientists, writes (p. S49): "In place of theoretical criteria, various prediction competitions have been used to grade algorithms in the so-called 'Common Task Framework.'. . . None of this is a good substitute for a so-far nonexistent theory of optimal prediction."

Efron is correct that prediction competitions are not a satisfactory way to evaluate prediction methods. However, he is not correct when he states that a theory of optimal prediction is "so-far nonexistent." Wald (1939, 1945, 1950) considered the general problem of using sample data to make decisions. He posed the task as choice of a *statistical decision function*, which maps potentially available data into a choice among the feasible actions. His development of statistical decision theory provides a broad framework for decision making with sample data, yielding optimal decisions when these are well-defined and proposing criteria for "reasonable" decision making more generally.

Wald recommended ex ante (frequentist) evaluation of statistical decision functions as *procedures* applied as the sampling process is engaged repeatedly to draw independent data samples. Whereas computer scientists measure performance when a prediction method is trained on one sample and used to predict outcomes in a test sample, statistical decision theory measures average performance across all possible training samples, when the objective is to predict outcomes in an entire population rather than a test sample.

The idea of a procedure transforms the inductive problem of evaluating a prediction method based on its performance in a single setting into the deductive problem of assessing the performance of a statistical decision function across realizations of the sampling process. It enables coherent study of treatment choice using sample data to make probabilistic predictions, with application to personalized medicine and elsewhere. This paper shows how.

The findings reported here add to a recent econometric literature using statistical decision theory to study treatment choice with sample data. See Manski (2004, 2005, 2007b, 2019b, 2021), Manski and Tetenov (2007, 2016, 2019, 2021), Hirano and Porter (2009, 2020), Stoye (2009, 2012), Tetenov (2012), Kitagawa and Tetenov (2018), Mbakop and Tabord-Meehan (2021), and Athey and Wager (2021).

Relative to this precedent work, part of the contribution of the present paper is its consideration of a class of treatment-choice problems that differs in some respects from those studied earlier. Part is its new application of the theme of *as-if optimization*, developed abstractly in Manski (2021). Part is its cautionary advice to clinical researchers and clinicians as they seek to interpret personalized medical risk assessments evaluated using traditional biostatistical criteria or prediction competitions.

Section 2 explains in broad terms how statistical decision theory enables study of treatment choice using sample data to make probabilistic predictions. I draw on and extend exposition in Manski (2021). Section 3 explains use of as-if optimization to choose between surveillance and aggressive treatment, first in generality and then when the data are generated by random sampling from P(y|x). Section 4 studies as-if optimization using estimates of P(y|x) that combine data on persons with different covariate values. A particular innovation is to introduce a new form of analysis of kernel estimation. Section 5 concludes.

## 2. Statistical Decision Theory for Binary Treatment Choice

### 2.1. Basic Elements of Statistical Decision Theory

Wald began with the standard decision theoretic problem of a planner who must choose an action yielding welfare that depends on an unknown state of nature. The planner specifies a state space listing the states considered possible. He chooses without knowing the true state. Wald added to this standard problem by supposing that the planner observes sample data that may be informative about the true state.

In the context of this paper, the action is a treatment choice, the unknown state of nature is the conditional probability distribution $P(y|x)$, and the sample data are informative about $P(y|x)$. I describe basic ideas in abstraction before applying them to this context.

### 2.1.1. Decisions without Sample Data

First consider decisions without sample data. A planner faces a choice set C and believes that the true state of nature $s^*$ lies in state space S. An objective function $w(\cdot, \cdot): C \times S \to R^1$ maps actions and states into welfare. The planner ideally would maximize $w(\cdot, s^*)$ over C, but he does not know $s^*$. To choose an action, decision theorists have proposed various ways of using $w(\cdot, \cdot)$ to form functions of actions alone, which can be optimized. When posing extremum problems, I use max and min notation, without concern for the subtleties that sometimes make it necessary to use sup and inf operations.

One approach places a subjective probability distribution $\pi$ on the state space, computes average state-dependent welfare with respect to $\pi$, and maximizes subjective average welfare over C. The criterion solves

$$(1) \quad \max_{c \,\in\, C} \ \int w(c, s) d\pi.$$

Another approach seeks an action that, in some sense, works uniformly well over all of S. This yields the maximin and minimax-regret (MMR) criteria. The maximin criterion solves the problem

$$(2) \quad \max_{c \,\in\, C} \ \min_{s \,\in\, S} \ w(c, s).$$

The MMR criterion solves

(3)     $\min\limits_{c \in C} \quad \max\limits_{s \in S} \quad [\max\limits_{d \in C} w(d, s) - w(c, s)]$.

Here $\max_{d \in C} w(d, s) - w(c, s)$ is the *regret* of action c in state s. The true state being unknown, one evaluates c by its maximum regret over all states and selects an action that minimizes maximum regret. The maximum regret of an action measures its maximum distance from optimality across states.

### 2.1.2. Statistical Decision Problems

Statistical decision problems suppose that the planner observes data generated by a sampling distribution. Knowledge of the sampling distribution is generally incomplete. To express this, one extends state space S to list the feasible sampling distributions, denoted ($Q_s$, s ∈ S). Let $\Psi_s$ denote the sample space in state s; $\Psi_s$ is the set of samples that may be drawn under distribution $Q_s$. The literature typically assumes that the sample space does not vary with s and is known. I assume this and denote the sample space as $\Psi$. Then a statistical decision function (SDF), c(·): $\Psi \rightarrow C$, maps the sample data into a chosen action.

An SDF is a deterministic function after realization of the sample data, but it is a random function ex ante. Hence, the welfare achieved is a random variable ex ante. Wald's theory evaluates the performance of SDF c(·) in state s by $Q_s\{w[c(\psi), s]\}$, the ex-ante distribution of welfare that it yields across realizations $\psi$ of the sampling process.

It remains to ask how a planner might compare the welfare distributions yielded by different SDFs. Wald proposed measurement of the performance of c(·) in state s by its expected welfare across samples; that is, $E_s\{w[c(\psi), s]\} \equiv \int w[c(\psi), s]dQ_s$. Not knowing the true state, a planner evaluates c(·) by the expected welfare vector $(E_s\{w[c(\psi), s]\}$, s ∈ S).

Statistical decision theory has mainly studied the same decision criteria as has decision theory without sample data. Let $\Gamma$ be a specified set of SDFs, each mapping $\Psi \rightarrow C$. The statistical versions of criteria (1), (2), and (3) are

(4)     $\max_{c(\cdot) \in \Gamma} \int E_s\{w[c(\psi), s]\} \, d\pi,$

(5)     $\max_{c(\cdot) \in \Gamma} \min_{s \in S} E_s\{w[c(\psi), s]\},$

(6)     $\min_{c(\cdot) \in \Gamma} \max_{s \in S} (\max_{d \in C} w(d, s) - E_s\{w[c(\psi), s]\}).$

Observe that these ex-ante criteria for evaluation of performance differ fundamentally from the computer-science practice of ex-post evaluation of predictions on test samples. The Wald framework evaluates a decision criterion by its mean performance across all feasible samples, not by its performance in a particular sample. In the computer-science approach, some data collection process generates two datasets, say $\psi_{tr}$ and $\psi_{test}$, denoted the training and test samples. The training sample is used to compute a predictor function, say $pred(\psi_{tr})$. The predictor function is applied to the test sample, the test being how well it predicts some feature of $\psi_{test}$ deemed to be of interest. Judgement of how well the predictor function performs is typically subjective rather than through use of a formal statistical criterion.[1] Computer scientists often motivate their approach by stating that it protects against drawing misleading conclusions from prediction performance on the study sample, which may be unrealistically high due to so-called "overfitting" the data. Concern with overfitting does not arise in the Wald framework, which evaluates performance across all feasible samples, not in a particular study sample.

---

[1] The possibility of formal statistical analysis depends on how training and test samples are generated. In some cases, a well-understood sampling process generates data, which are then randomly split into training and testing subsamples. In other cases, an idiosyncratic process generates data, which are then randomly split as above. In yet other cases, separate idiosyncratic processes generate training and test samples. In principle, it should be possible to study cases of the first type using classical frequentist statistical theory. In cases of the second type, the initial data generation is not interpretable with statistical theory, but the randomized split into training and test samples may enable randomization inference. Cases of the third type may not be amenable to any formal statistical analysis.

Manski (2004, 2021) discuss and compare the properties of criteria (4) – (6). To summarize some main points, maximization of subjective average welfare (4) may be appealing if one has a credible basis to place a subjective probability distribution on the state space, but not otherwise. Concern with specification of priors motivated Wald to study the maximin criterion (5). However, I see conceptual reasons to focus instead on the MMR criterion (6).

The conceptual appeal of using maximum regret to measure performance is that it quantifies how lack of knowledge of the true state of nature diminishes the quality of decisions. The term "maximum regret" is a shorthand for the maximum sub-optimality of a decision criterion across the feasible states of nature. An SDF with small maximum regret is uniformly near-optimal across all states. This is a desirable property.

Subject to regularity conditions ensuring that the relevant expectations and extrema exist, problems (4) – (6) offer criteria for decision making with sample data that are broadly applicable in principle. The primary challenge is computational. Problems (4) − (6) have tractable analytical solutions only in certain cases. Computation commonly requires numerical methods to find approximate solutions.

Expected welfare $E_s\{w[c(\psi), s]\}$ typically does not have an explicit form, but it can be well-approximated by Monte Carlo integration. One draws repeated values of $\psi$ from distribution $Q_s$, computes the average value of $w[c(\psi), s]$ across the values drawn, and uses this to estimate $E_s\{w[c(\psi), s]\}$. Monte Carlo integration can also be used in criterion (4) to approximate the subjective average of expected welfare.

The main computational challenges are determination of the extrema across actions in problem (6), across states in problems (5) − (6), and across SDFs in problems (4) − (6). Solution of $\max_{d \in C} w(d, s)$ in (6) is often straightforward but sometimes difficult. Finding extrema over S must cope with the fact that the state space commonly is uncountable. In applications where the quantity to be optimized varies smoothly over S, a simple approach is to compute the extremum over a suitable finite grid of states.

The most difficult computational challenge usually is to optimize over the feasible SDFs, $\Gamma$. No generally applicable approach is available. Hence, applications of statistical decision theory necessarily proceed case-by-case. It may not be tractable to find the best feasible SDF, but one often can evaluate the performance of relatively simple SDFs that researchers use in practice. This paper studies illustrative cases.

## 2.2. Binary Choice Problems

SDFs for binary choice have a simple structure. Manski (2021) shows that they can be viewed as hypothesis tests. Yet the Wald perspective on testing differs from the classical perspective of Neyman and Pearson (1928, 1933).

Let the choice set be C = {A, B}. An SDF $c(\cdot)$ partitions $\Psi$ into two regions that separate the data yielding choice of each action. These are $\Psi_{c(\cdot)A} \equiv [\psi \in \Psi: c(\psi) = A]$ and $\Psi_{c(\cdot)B} \equiv [\psi \in \Psi: c(\psi) = B]$. A test motivated by the choice problem partitions S into two regions, $S_A$ and $S_B$, that separate the states in which actions A and B are uniquely optimal. Thus, $S_A$ contains the states $[s \in S: w(A, s) > w(B, s)]$ and $S_B$ contains $[s \in S: w(B, s) > w(A, s)]$. The choice problem does not provide a rationale to allocate states where the actions yield equal welfare. A standard practice gives one action, say A, a privileged status and places states yielding equal welfare in $S_a$. Then $S_A \equiv [s \in S: w(A, s) \geq w(B, s)]$ and $S_B \equiv [s \in S: w(B, s) > w(A, s)]$.

In the language of testing, SDF $c(\cdot)$ performs a test with acceptance regions $\Psi_{c(\cdot)A}$ and $\Psi_{c(\cdot)B}$. When $\psi \in \Psi_{c(\cdot)A}$, $c(\cdot)$ accepts hypothesis $\{s \in S_A\}$ by setting $c(\psi) = A$. When $\psi \in \Psi_{c(\cdot)B}$, $c(\cdot)$ accepts $\{s \in S_B\}$ by setting $c(\psi) = B$. A test yields a Type I error when the true state lies in $S_A$ but $c(\psi) = B$. It yields a Type II error when the true state lies in $S_B$, but $c(\psi) = A$.

Although SDFs for binary choice are interpretable as tests, Neyman-Pearson testing and statistical decision theory evaluate tests differently. Neyman-Pearson testing views states $\{s \in S_A\}$ and $\{s \in S_B\}$ asymmetrically, calling the former the null hypothesis and the latter the alternative. A longstanding convention has been to restrict attention to tests in which the probability of a Type I error is no larger than a predetermined value, usually 0.05, for all $s \in S_a$. Then one restricts attention to SDFs $c(\cdot)$ for which $Q_s[c(\psi) = B] \leq 0.05$ for all $s \in S_a$.

Decision theory does not restrict attention to tests that yield a predetermined upper bound on the probability of a Type I error. Wald (1939) proposed evaluation of the performance of an SDF for binary choice by the expected welfare that it yields across realizations of the sampling process. The welfare distribution in state s in a binary choice problem is Bernoulli, with mass points max [w(a, s), w(b, s)] and min [w(a, s), w(b, s)]. These coincide if w(a, s) = w(b, s). When w(a, s) ≠ w(b, s), let $R_{c(\cdot)s}$ denote the probability that c(·) yields an error, choosing the inferior action over the superior one. That is,

(7)  $R_{c(\cdot)s}$  =  $Q_s[c(\psi) = b]$   if w(a, s) > w(b, s),

=  $Q_s[c(\psi) = a]$   if w(b, s) > w(a, s).

These are the probabilities of Type I and Type II errors.

The probabilities that welfare equals max [w(a, s), w(b, s)] and min [w(a, s), w(b, s)] are $1 - R_{c(\cdot)s}$ and $R_{c(\cdot)s}$. Hence, expected welfare is

(8)  $E_s\{w[c(\psi), s]\}$  =  $R_{c(\cdot)s}\{\min [w(a, s), w(b, s)]\} + [1 - R_{c(\cdot)s}]\{\max [w(a, s), w(b, s)]\}$

=  max [w(a, s), w(b, s)]  −  $R_{c(\cdot)s} \cdot |w(a, s) - w(b, s)|$.

Observe that $R_{c(\cdot)s} \cdot |w(a, s) - w(b, s)|$ is the expected regret of c(·). Thus expected regret, which was defined in abstraction in (6), has a simple form when choice is binary. It is the product of the error probability and the magnitude of the welfare loss when an error occurs.

### 2.3. As-If Optimization

The concept of a statistical decision function embraces all mappings [data → action]. An SDF need not perform inference; that is, it need not use data to draw conclusions about the true state of nature.

Although SDFs need not perform inference, some do. These have the form [data → inference → action], first performing inference and then using the inference to make a decision. There has been no accepted term for such SDFs, so Manski (2021) calls them *inference-based*.

A common type of inference-based SDF performs *as-if optimization*, also called *plug-in* or *two-step* decision making, choosing an action that optimizes welfare as if an estimate of the true state of nature actually is the true state. Formally, a point estimate is a function $s(\cdot)$: $\Psi \twoheadrightarrow S$ that maps data into the state space. As-if optimization means solving the problem $\max_{c \in C} w[c, s(\psi)]$. The result is an SDF $c[s(\cdot)]$, where

$$(9) \quad c[s(\psi)] \in \underset{c \in C}{\text{argmax}} \ w[c, s(\psi)], \quad \psi \in \Psi.$$

Traditionally, researchers have given computational and asymptotic statistical rationales for acting in the manner of (9). Computationally, using a point estimate to maximize welfare is easier than solving problems (4) to (6). To further motivate as-if optimization, statisticians and econometricians cite limit theorems of asymptotic theory. They hypothesize a sequence of sampling processes indexed by sample size and a corresponding sequence of estimates. They show that the sequence is consistent when specified assumptions hold. They may also derive the rate of convergence and limiting distribution of the estimate.

Computational and asymptotic arguments do not prove that as-if optimization provides a well-performing SDF. Statistical decision theory evaluates as-if optimization in state s by the expected welfare, $E_s\{w\{c[s(\psi)], s\}\}$, that it yields across samples of specified size, not asymptotically. This is how I proceed below when studying treatment choice with sample data.

## 3. Choice Between Surveillance and Aggressive Treatment

I now use statistical decision theory to study a version of the medical problem of choice between surveillance and aggressive treatment. The broad problem concerns a clinician caring for patients with observed covariates x. There are two care options for a specified disease, with A denoting surveillance and B denoting aggressive treatment. The clinician must choose without knowing a patient's illness status; y = 1 if a patient is ill and y = 0 if not. Observing x, the clinician can attempt to learn the conditional probability of illness, $p_x \equiv P(y = 1|x)$. I suppose that the planner performs as-if optimization, using sample data to estimate $p_x$ and acting as if the estimate is correct.

The version of the decision problem studied here maintains simplifying assumptions used in parts of the analysis in Manski (2018, 2019a). I assume that patient welfare with care option $c \in \{A, B\}$ has the known form $U_x(c, y)$; thus, welfare may vary with whether the disease occurs and with the patient covariates x. Aggressive treatment is better if the disease occurs, and surveillance is better otherwise. That is,

(10a)    $U_x(B, 1) > U_x(A, 1)$,

(10b)    $U_x(A, 0) > U_x(B, 0)$.

The specific form of welfare function $U_x(\cdot, \cdot)$ necessarily depends on the clinical context, but inequalities (10a) – (10b) are realistic in many settings.

I assume that the chosen care option does not affect whether the disease occurs; hence, a patient's illness probability is simply $p_x$ rather than a function $p_x(c)$ of the care option. With this assumption, treatment choice still matters because it may affect the severity of illness and patient experience of side effects. Aggressive treatment if beneficial to the extent that it lessens the severity of illness, but harmful if it yields side effects that do not occur with surveillance.

*Illustration*: A patient presents to a clinician with symptoms of a sore throat. The patient may have a streptococcal infection (y = 1) or a throat irritation (y = 0). Treatment A is to counsel the patient to rest and monitor body temperature until the result of a throat culture is obtained. Treatment B is immediate prescription of an antibiotic. The antibiotic will lessen the severity of illness if $y = 1$, but it will have no beneficial effect if $y = 0$. Whether or not infection is present, the patient may suffer an adverse side effect from receipt of the antibiotic. ∎

The central difference between Manski (2018, 2019a) and the present paper is that I earlier supposed the clinician may have deterministic partial knowledge of $p_x$ but has no informative sample data. Here I study treatment choice using sample data to estimate $p_x$. In the first part of this section, the estimate is an abstract function of sample data. This makes it easy to explain general principles. I subsequently specialize to settings where the data are drawn by random sampling of y conditional on x.

## 3.1. Treatment Choice with Knowledge of $p_x$

### 3.1.1. Optimal Treatment Choice

Before considering decision making with sample data, suppose that the clinician knows $p_x$ and chooses a treatment that maximizes expected patient welfare conditional on x. Then an optimal decision is

(11a)    Choose A  if  $p_x \cdot U_x(A, 1) + (1 - p_x) \cdot U_x(A, 0) \geq p_x \cdot U_x(B, 1) + (1 - p_x) \cdot U_x(B, 0)$,

(11b)    Choose B  if  $p_x \cdot U_x(B, 1) + (1 - p_x) \cdot U_x(B, 0) \geq p_x \cdot U_x(A, 1) + (1 - p_x) \cdot U_x(A, 0)$.

The decision yields optimal expected patient welfare

(12)      $\max [p_x \cdot U_x(A, 1) + (1 - p_x) \cdot U_x(A, 0), \, p_x \cdot U_x(B, 1) + (1 - p_x) \cdot U_x(B, 0)]$.

The optimal decision is easy to characterize when inequalities (10a) − (10b) hold. Let $p_x^*$ denote the threshold value of $p_x$ that makes options A and B have the same expected utility. This value is

$$(13) \qquad p_x^* = \frac{U_x(A, 0) - U_x(B, 0)}{[U_x(A, 0) - U_x(B, 0)] + [U_x(B, 1) - U_x(A, 1)]} .$$

Observe that $0 < p_x^* < 1$. Option A is optimal if $p_x \leq p_x^*$ and B if $p_x \geq p_x^*$. Thus, optimal treatment choice does not require exact knowledge of $p_x$. It only requires knowing whether $p_x$ is larger or smaller than $p_x^*$.

### 3.1.2. Aggressive Treatment Neutralizes Disease

An instructive special case occurs when aggressive treatment neutralizes disease, in the sense that $U_x(B, 0) = U_x(B, 1)$. For example, aggressive treatment might be surgery to remove a localized tumor that may ($y = 1$) or may not ($y = 0$) be malignant. Suppose that surgery always eliminates cancer when present. Then surgery neutralizes disease. Being invasive and costly, performance of surgery has a negative side effect on welfare that is the same regardless of whether cancer is present.

Let $U_{xB}$ denote welfare with aggressive treatment. Then (10) – (13) reduce to

$$(14) \qquad U_x(A, 0) > U_{xB} > U_x(A, 1).$$

$$(15a) \quad \text{Choose A if } p_x \cdot U_x(A, 1) + (1 - p_x) \cdot U_x(A, 0) \ \geq \ U_{xB},$$

$$(15b) \quad \text{Choose B if } U_{xB} \ \geq \ p_x \cdot U_x(A, 1) + (1 - p_x) \cdot U_x(A, 0).$$

$$(16) \qquad \max \ [p_x \cdot U_x(A, 1) + (1 - p_x) \cdot U_x(A, 0), \ U_{xB}].$$

$$(17) \qquad p_x^* = \frac{U_x(A, 0) - U_{xB}}{U_x(A, 0) - U_x(A, 1)}.$$

Further simplification occurs when one normalizes the location and scale of welfare by setting $U_x(A, 0) = 1$ and $U_x(A, 1) = 0$. Then $(14) - (17)$ become

(18)    $1 > U_{xB} > 0.$

(19a)    Choose A  if  $1 - p_x \geq U_{xB},$

(19b)    Choose B  if  $U_{xB} \geq 1 - p_x.$

(20)    max $(1 - p_x, U_{xB}).$

(21)    $p_x^* = 1 - U_{xB}.$

I henceforth assume that aggressive treatment neutralizes disease and I normalize the welfare of surveillance as above.

## 3.2. Maximum Regret of As-If Optimization

Now suppose that the clinician does not know $p_x$. The state space lists all feasible values of $p_x$. With no knowledge, $(p_{sx}, s \in S) = [0, 1]$. With partial knowledge, S is a non-singleton proper subset of $[0, 1]$. In the abstract notation of Section 2, the objective function $w(\cdot, \cdot): C \times S \rightarrow R^1$ has this form: $w(A, s) = p_{sx} \cdot U_x(A, 1) + (1 - p_{sx}) \cdot U_x(A, 0)$  and $w(B, s) = U_{xB}.$

I assume that the clinician does not know whether $p_x$ is smaller or larger than $p_x^* = 1 - U_{xB}$. Formally, $p_{mx} < 1 - U_{xB} < p_{Mx}$, where $p_{mx} \equiv \min_{s \in S} p_{sx}$ and $p_{Mx} \equiv \max_{s \in S} p_{sx}$. Thus, the clinician cannot maximize expected patient welfare conditional on x.

The clinician uses sample data to estimate $p_x$ and acts as if the estimate is correct. Let $\Psi$ be a sample space and let $Q_s$ be a sampling distribution with realizations $\psi \in \Psi$. Let $\varphi_x(\psi)$ be a point estimate of $p_x$. The clinician maximizes expected welfare acting as if $\varphi_x(\psi) = p_x$. Thus, the chosen care option is A if $1 - \varphi_x(\psi) \geq U_{xB}$ and is B if $U_{xB} > 1 - \varphi_x(\psi)$. I analyze treatment choice from the perspective of maximum regret.

Let $e[p_{sx}, \varphi_x(\psi), U_{xB}]$ denote the occurrence of an error in state s when $\varphi_x(\psi)$ is used to choose treatment. That is, $e[p_{sx}, \varphi_x(\psi), U_{xB}] = 1$ when $p_{sx}$ and $\varphi_x(\psi)$ yield different treatments, while $e[p_{sx}, \varphi_x(\psi), U_{xB}] = 0$ when $p_{sx}$ and $\varphi_x(\psi)$ yield the same treatment. Regret using estimate $\varphi_x(\psi)$ is

$$
\begin{aligned}
(22) \quad R_{sx}[\varphi_x(\psi)] \;&=\; \max\,(1 - p_{sx},\, U_{xB}) \,-\, (1 - p_{sx}){\cdot}1[1 - \varphi_x(\psi) \geq U_{xB}] \,-\, U_{xB}{\cdot}1[U_{xB} > 1 - \varphi_x(\psi)] \\
&=\; |(1 - p_{sx}) - U_{xB}|{\cdot}1[1 - p_{sx} \geq U_{xB} > 1 - \varphi_x(\psi) \text{ or } 1 - p_{sx} < U_{xB} \leq 1 - \varphi_x(\psi)] \\
&=\; |(1 - p_{sx}) - U_{xB}|{\cdot}e[p_{sx}, \varphi_x(\psi), U_{xB}].
\end{aligned}
$$

Expected regret across repeated samples is

$$
(23) \quad E_s\{R_{sx}[\varphi_x(\psi)]\} \;=\; |(1 - p_{sx}) - U_{xB}|{\cdot}Q_s\{e[p_{sx}, \varphi_x(\psi), U_{xB}] = 1\}.
$$

Maximum expected regret across the state space is $\max_{s \in S} E_s\{R_{sx}[\varphi_x(\psi)]\}$.

Evaluation of estimate $\varphi_x(\cdot)$ by the maximum regret of the treatment choices it yields is reminiscent of, but distinct in various respects from, analysis of plug-in decision making in the statistical-learning literature on *pattern recognition* or *classification* (e.g., Devroye *et al.,* 1996; Yang, 1999). Interpreted in the context of the present paper, researchers in that field have supposed that the welfare function is anti-symmetric, with $[U_x(A, 0) = c, U_x(B, 0) = d]$ and $[U_x(A, 1) = d, U_x(B, 1) = c]$, for some $c > d$. Then the threshold for decision making is $p_x^* = \frac{1}{2}$. They have assumed that the data are a random sample of persons from a population with heterogeneous covariate values. They have sought to measure the performance of a decision criterion by maximum regret when the criterion is applied across the population rather than to persons with a specified covariate value. They have studied asymptotic questions of consistency and rates of convergence: does maximum regret converge to zero as sample size increases and, if so, how fast? As far as I am aware, they have not studied computation of finite-sample maximum regret, which is the focus of this paper.[2]

Evaluation of $\varphi_x(\cdot)$ by maximum regret differs fundamentally from computer-science evaluation by ex-post prediction accuracy. In that paradigm, one uses a training sample $\psi_{tr}$ to compute $\varphi_x(\psi_{tr})$ and then uses this estimate to predict illness in a test sample $\psi_{test}$. A standard practice is to predict y = 1 if $\varphi_x(\psi_{tr})$ exceeds a specified threshold, say $\gamma$, and y = 0 otherwise. Prediction accuracy is measured by the *positive predictive value*, the fraction of the test sample with y = 1 conditional on $\varphi_x(\psi_{tr}) \geq \gamma$, and the *negative predictive value*, the fraction of the test sample with y = 0 conditional on $\varphi_x(\psi_{tr}) \leq \gamma$. These measures of ex post prediction accuracy do not consider performance across samples or over the state space. Nor do they consider the patient welfare achieved when an estimate is used to choose a treatment.

### 3.2.1. Computation of Maximum Regret

Numerical computation of maximum regret with as-if optimization is generally tractable. The error probability $Q_s\{e[p_{sx}, \varphi_x(\psi), U_{xB}] = 1\}$ can be approximated by Monte Carlo integration. One draws repeated values of $\psi$ from distribution $Q_s$. One computes the fraction of cases in which the values drawn generate estimates that yield errors in treatment choice. One uses this fraction to estimate the error probability. The statistical precision of the estimate of the error probability increases with the number of $\psi$ values drawn.

---

2 Another form of as-if optimization has been studied in econometrics and statistical learning theory. Rather than estimate $p_x$ and then determine whether the estimate exceeds the threshold $p_x^*$, one directly estimates the indicator function $1[p_x > p_x^*]$ and makes a decision accordingly. When the data are a random sample from a population with heterogeneous covariates, this approach yields the *maximum score* method of econometrics (Manski, 1975, 1985; Manski and Thompson, 1989) and the *empirical risk minimization* methods of statistical learning theory (Vapnik, 1999, 2000). Here too, researchers have mainly studied asymptotic questions of consistency and rates of convergence.

Expected regret is easy to compute; it equals the error probability times $|(1 - p_{sx}) - U_{xB}|$. Maximizing regret must cope with the fact that the set ($p_{sx}$, $s \in S$) commonly is uncountable. Being a subset of [0, 1], this set is relatively simple in structure. A pragmatic approach is to maximize over a suitable finite grid of feasible probability values. Refining the grid increases the accuracy of the approximate solution.

Computation is particularly straightforward when the data are illness outcomes ($y_i$, $i = 1, \ldots, N_x$) that have been observed in a random sample of $N_x$ persons drawn from a study population with illness probability $p_x$. The ordering of the observations in a random sample is immaterial, so the sample space may be defined to be the number $n_x$ of observed illness outcomes; thus, $\Psi = \{0, 1, 2, \ldots, N_x\}$. The sampling distribution in state s is the Binomial distribution $Q_s = \mathbf{B}(p_{sx}, N_x)$, where

$$(24) \qquad f(n_x; p_{sx}, N_x) \equiv N_x! [n_x! \cdot (N_x - n_x)!]^{-1} p_{sx}^{n_x} (1 - p_{sx})^{N_x - n_x}$$

is the probability of observing $n_x$ illnesses.

In this setting, expected regret has the form

$$(25a) \quad E_s\{R_{sx}[\varphi_x(\psi)]\} \; = \; [(1 - p_{sx}) - U_{xB}] \cdot \mathbf{B}[U_{xB} > 1 - \varphi_x(n); p_{sx}, N_x] \; \text{ for } s \in S_A,$$

$$(25b) \quad E_s\{R_{sx}[\varphi_x(\psi)]\} \; = \; [U_{xB} - (1 - p_{sx})] \cdot \mathbf{B}[U_{xB} \leq 1 - \varphi_x(n); p_{sx}, N_x] \; \text{ for } s \in S_B,$$

where $S_A = (s \in S: 1 - p_{sx} \geq U_{xB})$ and $S_B = (s \in S: 1 - p_{sx} < U_{xB})$ were defined in Section 2.2. Computing maximum regret when performing as-if optimization with estimate $\varphi_x(\cdot)$ requires maximizing (25a) over $s \in S_A$ and maximizing (25b) over $s \in S_B$. Maximum regret over S is the larger of these sub-maxima. The sub-maximization problems usually do not have explicit solutions. However, the fact that n has the Binomial distribution makes it easy to perform numerical maximization. Illustrative findings are given in Section 4.

While numerical computation of maximum regret must be the norm, extreme cases with no informative sample data or with one observation of y drawn at random from $p_x$ are amenable to simple analysis. An analytical upper bound on maximum regret is available when multiple observations of y are drawn at random from $p_x$ and the sample illness frequency is used to estimate $p_x$. Sections 3.3 through 3.6 present the findings.

## 3.3. Minimax Regret with No Data

Suppose that one observes no sample data. Then the estimate must be a constant $\varphi_x$ rather than a random variable $\varphi_x(\psi)$. Under the maintained assumption that $p_{mx} \equiv \min_{s \in S} p_{sx} < 1 - U_{xB} < p_{Mx} \equiv \max_{s \in S} p_{sx}$, the minimax regret estimate is given in Proposition 1. The proofs of this and all subsequent propositions are collected in an Appendix.

<u>Proposition 1</u>: Consider as-if optimization with no sample data, using an estimate $\varphi_x$. Estimates such that $U_{xB} > 1 - \varphi_x$ minimize maximum regret if $(1 - p_{mx}) - U_{xB} \leq U_{xB} - (1 - p_{Mx})$. Estimates such that $U_{xB} \leq 1 - \varphi_x$ minimize maximum regret if $(1 - p_{mx}) - U_{xB} \geq U_{xB} - (1 - p_{Mx})$. The minimum achievable value of maximum regret is

(26)     $\min[(1 - p_{mx}) - U_{xB}, \; U_{xB} - (1 - p_{Mx})]$.     ∎

The MMR results simplify if the clinician has no prior knowledge, so $p_{mx} = 0$ and $p_{Mx} = 1$. Then $\varphi_x$ such that $U_{xB} > 1 - \varphi_x$ minimize maximum regret if $1 - U_{xB} \leq U_{xB}$, whereas $\varphi_x$ such that $U_{xB} \leq 1 - \varphi_x$ minimize maximum regret if $1 - U_{xB} \geq U_{xB}$. Thus, $\varphi_x$ such that $U_{xB} > 1 - \varphi_x$ minimize maximum regret if $\frac{1}{2} \leq U_{xB}$, and $\varphi_x$ such that $U_{xB} \leq 1 - \varphi_x$ minimize maximum regret if $\frac{1}{2} \geq U_{xB}$. The minimum achievable value of maximum regret is $\min(1 - U_{xB}, U_{xB})$.

## 3.4. Minimax Regret with Uninformative Sample Data

Suppose that one does not observe data that are informative about the true state of nature $s^*$. Nevertheless, it is always possible to generate random data that are uninformative about $s^*$ and use them to estimate $p_x$. One may specify a distribution on the interval $[0, 1]$ and estimate $p_x$ by a realization drawn from this distribution.

As-if optimization with an estimate based on uninformative data may seem pointless, but it opens new possibilities for treatment choice relative to the situation with no data at all. Estimates were deterministic in that case, so error probabilities could take only the value 0 or 1. Now estimates can be random variables and error probabilities can take any value in $[0, 1]$. I show that this makes it possible to reduce maximum regret. The present analysis is broadly similar to my earlier work (Manski, 2009, 2021) showing that randomization improves on deterministic binary treatment choice, but it differs in the details.

Formally, one specifies a sample space $\Psi_0$, a sampling distribution $Q_0$ on $\Psi_0$, and an estimate $\varphi_{0x}(\cdot)$: $\Psi_0 \to [0, 1]$. One programs a random number generator to draw realizations $\psi$ with distribution $Q_0$, and one uses $\varphi_{0x}(\psi)$ to estimate $p_x$. The minimax regret estimate is given in Proposition 2.

<u>Proposition 2</u>: Consider as-if optimization with uninformative sample data, using an estimate $\varphi_{0x}(\cdot)$. Let $q_{0x} \equiv Q_0[U_{xB} > 1 - \varphi_{0x}(\psi)]$. Estimates such that

$$(27) \quad q_{0x} = \frac{U_{xB} - (1 - p_{Mx})}{p_{Mx} - p_{mx}}$$

minimize maximum regret. The minimum achievable value of maximum regret is

$$(28) \quad \frac{[(1 - p_{mx}) - U_{xB}] \cdot [U_{xB} - (1 - p_{Mx})]}{p_{Mx} - p_{mx}} . \qquad \blacksquare$$

Proposition 1 showed that minimum achievable maximum regret using no sample data is (26). Expression (28) is smaller than (26) under the maintained assumption that $p_{mx} < 1 - U_{xB} < p_{Mx}$. Thus, as-if optimization with uninformative sample data reduces minimum achievable maximum regret relative to treatment choice with no sample data.

3.5. Minimax Regret with One Illness Observation Drawn at Random from $p_x$

Here and in Section 3.6, I suppose that one observes $N_x$ illness outcomes drawn at random from $p_x$. It is then natural to consider as-if optimization with estimate $\varphi_x(n_x) = n_x/N_x$, which uses the sample rate of illness to estimate the illness probability. Proposition 3 gives the exact value of maximum regret when $N_x = 1$ and the state space is $S = [0, 1]$. Section 3.6 derives an upper bound on maximum regret that holds for any specified value of $N_x$.

Proposition 3: Let $S = [0, 1]$. Let the sample data be one illness observation drawn at random from $p_x$. Consider as-if optimization with estimate $[\varphi_x(0) = 0, \varphi_x(1) = 1]$. The value of maximum regret is

$$(29) \quad \max_{s \in S} E_s\{R_{sx}[\varphi_x(\psi)]\} = \tfrac{1}{4} \max[(1 - U_{xB})^2, U_{xB}^2]. \qquad \blacksquare$$

The proposition shows that as-if optimization with estimate $[\varphi_x(0) = 0, \varphi_x(1) = 1]$ substantially outperforms treatment choice with data-invariant estimates or uninformative data when $U_{xB} = \tfrac{1}{2}$. Using the sample data yields maximum regret 1/16. Using no data or uninformative data yield maximum regret $\tfrac{1}{2}$ or $\tfrac{1}{4}$ respectively. However, as $U_{xB}$ moves away from $\tfrac{1}{2}$, performance with estimate $[\varphi_x(0) = 0, \varphi_x(1) = 1]$ deteriorates and performance not using the data improves. Indeed, as-if optimization with the data-invariant estimates [$\varphi_x(0) = 0$, $\varphi_x(1) = 0$] and [$\varphi_x(0) = 1$, $\varphi_x(1) = 1$] have smaller maximum regret when $U_{xB} < 3 - 2\sqrt{2}$ and $U_{xB} > 2(\sqrt{2} - 1)$ respectively.

### 3.6. Upper Bound on Maximum Regret Using the Sample Illness Rate to Estimate $p_x$

Now suppose that one observes $N_x \geq 1$ illness outcomes drawn at random from $p_x$ and performs as-if optimization with estimate $\varphi_x(n_x) = n_x/N_x$. Asymptotic theory suggests this approach to treatment choice. The Strong Law of Large Numbers implies that $n_x/N_x \to p_x$ almost surely as $N_x \to \infty$.

An analytical finite-sample justification stems from a large-deviations inequality of Hoeffding (1963) for averages of bounded random variables, which shows that $Q_s(n_x/N_x - p_{sx} > \delta) \leq \exp(-2N_x\delta^2)$ and $Q_s(p_{sx} - n_x/N_x > \delta) \leq \exp(-2N_x\delta^2)$ for all $\delta > 0$. These inequalities yield an upper bound on maximum regret, whose magnitude depends on the known values of $(p_{mx}, p_{Mx}, U_{xB})$. Proposition 4 gives the bound.

<u>Proposition 4</u>: Consider as-if optimization with estimate $\varphi_x(n) = n_x/N_x$. Then, for all $\delta > 0$,

$$(30) \quad \max_{s \in S} \; E_s[R_{sx}(n_x/N_x)] \; \leq \; \delta \; + \; \{\max \; [(1 - p_{mx}) - U_{xB}, \; U_{xB} - (1 - p_{Mx})]\} \cdot \exp(-2N_x\delta^2). \quad \blacksquare$$

Minimizing the analytical bound (30) over $\delta > 0$ yields a tighter bound that can be determined numerically. Given any value of $(p_{mx}, p_{Mx}, U_{xB})$, the bound decreases to zero if $\delta \to 0$ and $N_x\delta^2 \to \infty$. Hence, the maximum regret of as-if optimization with estimate $\varphi_x(n) = n_x/N_x$ converges to zero as $N_x \to \infty$.

Bound (30) is simple and useful, but it is not sharp. Numerical computation of maximum regret is straightforward and shows that the exact value is sometimes much less than the bound. Hence, exact numerical computation is recommended in practice.

## 4. As-If Optimization Using Data on Persons with Different Covariates

Researchers often analyze data on outcomes for persons with heterogeneous covariates. When the decision problem is to choose a treatment for someone with a specific covariate value, data on persons with other covariates are not informative per se. However, these data may be informative when assumptions are imposed that relate the outcome distributions of persons with different covariates. An important subject for methodological research is to learn what is achievable with various combinations of data and assumptions. I demonstrate here, continuing to focus on maximum regret.

I focus on the instructive, simple setting where persons have either of two covariate values, $x = 0$ and $x = 1$. Persons with $x = 0$ and $x = 1$ may be similar in some respects, but they differ in some way. State $s$ now indexes a possible pair $(p_{s0}, p_{s1})$ of conditional illness probabilities. Let random samples of $N_0$ outcomes be drawn from $p_0$ and $N_1$ outcomes be drawn from $p_1$, these sample sizes being predetermined. Then the data are the numbers of ill persons in each sample, $n_0$ and $n_1$ respectively.[3]

Let the decision problem be to choose a treatment for a person with $x = 0$. The question of interest is the extent to which observation of $(n_0, n_1)$ improves treatment choice relative to observation of $n_0$ alone. Proposition 2 showed that observation of sample data can improve treatment choice even when the data are uninformative about $p_0$, because the data provide a means to randomize treatment choice. In this section, I consider settings in which one maintains assumptions that make observation of $n_1$ informative about $p_0$.

In the absence of assumptions that suitably restrict the state space, observation of $n_1$ is not informative about $p_0$. Under random sampling, the joint sampling distribution of $(n_0, n_1)$ in state $s$ is the Binomial product $\mathbf{B}(p_{s0}, N_0) \times \mathbf{B}(p_{s1}, N_1)$. The distribution of $n_1$ varies with the value of $p_{s1}$, but not with the value of $p_{s0}$. Hence, $n_1$ is uninformative about $p_0$.

---

[3] Considering $N_0$ and $N_1$ to be predetermined simplifies regret analysis relative to a setting where persons are sampled at random from the population at large. In that setting, $(N_0, N_1, n_0, n_1)$ are jointly random variables. With $(N_0, N_1)$ predetermined, only $(n_0, n_1)$ are random. Moreover, $n_0$ and $n_1$ are statistically independent of one another. The analysis performed here applies with sampling at random from the population at large if one measures performance by expected regret conditional on realized values of $(N_0, N_1)$ rather than by unconditional expected regret.

Observation of $n_1$ becomes informative when the state space has non-rectangular structure. A rectangular state space has the form $S = S_0 \times S_1$, where $S_0$ and $S_1$ index the feasible values of $p_0$ and $p_1$ respectively. Then the feasible values of $p_0$ do not vary with the value of $p_1$. If $S$ is non-rectangular, the feasible $p_0$ vary with $p_1$. Hence, observation of $n_1$ may be informative about $p_0$, via $p_1$. Sections 4.1 and 4.2 examine two settings with non-rectangular state spaces.

## 4.1. Bounded Variation Between Illness Probabilities

One may find it credible to assume that $p_0$ and $p_1$ are not too different from one another. Thus, one may impose a *bounded-variation* assumption of the form

$$(31) \quad p_{s1} + \lambda_- \leq p_{s0} \leq p_{s1} + \lambda_+, \quad \text{all } s \in S,$$

for specified $\lambda_- \leq \lambda_+$. The implications of bounded variation assumptions for identification of conditional probabilities have been studied by Manski and Pepper (2000, 2018) and Manski (2018). As far as I am aware, the only precedent work studying the implications for decision making with finite-sample data is Stoye (2012), who studied a class of treatment choice problems whose structure differs from the problem examined here.

*Illustration*: Manski (2018) examined how life span varies with (age, sex, race, hypertension status). Let x = 0 and x = 1 respectively denote black and white males of age 50 who have been diagnosed with hypertension. Let $p_x$ denote the conditional probability of death prior to age 70. The study conjectured that black males tend to face health disadvantages relative to white males beyond hypertension and, hence, that black males tend to have lower life spans than white males, conditional on current age and hypertension status. This yields the bound $p_1 \leq p_0$. Going further, one might find it credible that the probability of death prior to age 70 is at most a specified amount greater for blacks than whites, say 0.2. Then (31) holds with $\lambda_- = 0$ and $\lambda_+ = 0.2$.     ∎

### 4.1.1. Estimation Using the Combined Sample Average

One might use the combined sample average $(n_0 + n_1)/(N_0 + N_1)$ to estimate $p_0$. In the statistical literature, estimation by a combined average rather than by $n_0/N_0$ is called *dimension reduction*. Statisticians usually analyze dimension reduction as a tradeoff between variance and bias, the objective being to minimize the mean square error of prediction. Combining samples increases the total sample size from $N_0$ to $N_0 + N_1$, increasing precision. However, the quantity being estimated is now a weighted average of $p_0$ and $p_1$, which differs from $p_0$ if $p_1 \neq p_0$.

The intuition of a tradeoff between variance and bias extends to evaluation of maximum regret in binary treatment choice. However, maximum regret when using an estimate of $p_0$ in treatment choice differs from the maximum mean square error of the estimate. Hence, the mathematical analysis differs.

As-if optimization with estimate $(n_0 + n_1)/(N_0 + N_1)$ yields smaller maximum regret than using $n_0/N_0$ for some values of the parameters $(p_{m0}, p_{M0}, U_{0B}, N_0, N_1, \lambda_-, \lambda_+)$, but larger maximum regret for other values. Combining samples is obviously preferable when $\lambda_- = \lambda_+ = 0$, as (31) reduces to $p_{s1} = p_{s0}$ for all s. It also holds when $\lambda_-$ and $\lambda_+$ are not too far from zero. In these cases, the benefit of increasing sample size from $N_0$ to $(N_0 + N_1)$ exceeds the imperfection of using data on persons with illness probability $p_1$ to estimate illness probability $p_0$. Given specified values of the parameters, maximum regret using the two estimates can be computed numerically and compared. See Section 4.1.2.

As a prelude, Propositions 5 and 6 present analytical findings indicating when combining samples outperforms using $n_0/N_0$. To simplify analysis, these propositions assume that $p_{m1} = 0$, $p_{M1} = 1$, and that the bound in (31) is symmetric.

The first result concerns the special case where $N_0 = 0$ and $N_1 = 1$. When $N_0 = 0$, disregarding data for persons with $x = 1$ implies treatment choice with a data-invariant estimate. The estimate that uses the data is $n_1$. Proposition 5, which extends Proposition 3, gives maximum regret for as-if optimization using this estimate, provided that the bound on $p_0$ specified in (31) is symmetric and not too wide.

<u>Proposition 5</u>: Let $0 \le \lambda \le \min (U_{0B}, 1 - U_{0B})$. Let (31) hold, with $\lambda_+ = \lambda$ and $\lambda_- = -\lambda$. Let $p_{m1} = 0$ and $p_{M1} = 1$. Let the sample data be one realization drawn at random from $p_1$. Consider as-if optimization with estimate $n_1$. The value of maximum regret is

$$(32) \quad \max_{s \in S} E_s[R_{s0}(n_1)] \ = \ \tfrac{1}{4} \max[[(1 - U_{0B}) + \lambda]^2, (U_{0B} + \lambda)^2]. \qquad \blacksquare$$

Proposition 6, which extends Proposition 4, gives the second result.

<u>Proposition 6</u>: Let $0 \le \lambda$. Let (31) hold, with $\lambda_+ = \lambda$ and $\lambda_- = -\lambda$. Let $p_{m1} = 0$ and $p_{M1} = 1$. Consider as-if optimization with the estimate $(n_0 + n_1)/(N_0 + N_1)$. Let $\alpha_1 \equiv N_1/(N_0 + N_1)$. Then

$$(33) \quad \max_{s \in S} E_s\{R_{s0}[(n_0 + n_1)/(N_0 + N_1)]\} \ \le$$

$$(\delta + \alpha_1\lambda) + \{\max \, [(1 - p_{m0}) - U_{0B}, U_{0B} - (1 - p_{M0})]\}\cdot\exp[-2(N_0 + N_1)\delta^2]. \qquad \blacksquare$$

Comparison of bounds (33) and (30) shows that the first term of (33) exceeds that in (30), being $(\delta + \alpha_1\lambda)$ rather than $\delta$. However, the second term of (33) is less than that in (30), as $\exp[-2(N_0 + N_1)\delta^2]$ is less than $\exp(-2N_0\delta^2)$. Hence, the upper bound on maximum regret using estimate $(n_0 + n_1)/(N_0 + N_1)$ is smaller than the one using estimate $n_0/N_0$ when $\alpha_1\lambda$ is sufficiently small and $N_1$ is sufficiently large.

### 4.1.2. Estimation Using a Weighted Sample Average

The sample averages $(n_0 + n_1)/(N_0 + N_1)$ and $n_0/N_0$ provide polar ways to estimate $p_0$. The former acts as if $p_1 = p_0$, whereas the latter acts as if $p_1$ and $p_0$ may be arbitrarily different from one another. Between the two poles, one might consider estimation by a weighted average, data with $x = 0$ being weighted more heavily than data with $x = 1$. Such an estimate is

(34)  $\varphi_0(n_0, n_1) = (w_0 n_0 + w_1 n_1)/(w_0 N_0 + w_1 N_1),$

where $\frac{1}{2} \leq w_0 \leq 1$ and $w_1 = 1 - w_0$ are the weights. Weighted-average estimates perform partial dimension reduction, bridging the gap between complete dimension reduction ($w_0 = \frac{1}{2}$) and no reduction ($w_0 = 1$).

Equation (34) is a simple form of the kernel estimate studied in the statistical literature on nonparametric regression. However, maximum-regret analysis of the performance of the estimate when used in binary treatment choice differs considerably from standard analysis of kernel estimation. To the extent that statisticians have performed finite-sample analysis, the usual concern has been the maximum mean square of an estimate. The literature mainly studies asymptotic properties of estimates---convergence of mean square error to zero, convergence in probability, and rates of convergence as sample size increases. Theorems typically assume that x is a real vector whose distribution has positive density in a neighborhood of a value of interest. With some exceptions, theorems assume that the conditional expectation $E(y|x)$ varies smoothly with x in a local sense, such as being differentiable, rather than a global sense such as being Lipschitz or Hölder continuous. Thus, they typically do not impose bounded-variation assumptions such as (34), which bound the difference between $E(y|x = x_1)$ and $E(y|x = x_0)$ at specified covariate values. Donoho *et al.* (1995) reviews many findings.

Given specified values of $(p_{m0}, p_{M0}, U_{0B}, N_0, N_1, \lambda_-, \lambda_+, w_0)$, maximum regret using a weighted-average estimate can be computed numerically. Moreover, one can vary $w_0$ and determine the weighting that minimizes maximum regret among all weighted averages. To illustrate, I use the problem of treating bleeds in patients with immune thrombocytopenia (ITP).

*Illustration*: ITP is an autoimmune disease characterized by low platelet counts and increased risk of bleeding. When a patient with ITP presents in a hospital emergency department, a difficult clinical problem is to predict whether the patient is experiencing a *critical bleed* (y = 1) or not (y = 0).[4] A critical bleed warrants aggressive treatment, while surveillance is preferable otherwise.

Assume that aggressive treatment neutralizes disease by stopping a critical bleed, but it may have side effects whose implications for patient welfare are measured by $U_{xB}$. The treatment decision is made with knowledge of patient covariates x, but without knowledge of y. Given knowledge of $p_x$, the conditional probability that a bleed is critical, the optimal decision is surveillance if $1 - p_x \geq U_{xB}$ and aggressive treatment if $1 - p_x \leq U_{xB}$. Suppose that, $p_x$ not being known, treatment choice will be made by as-if optimization with a weighted average estimate.

For specificity, let x = 0 and x = 1 respectively denote female and male patients who have the same observed attributes other than gender. Suppose that available clinical knowledge and assessment of patient welfare makes it credible to set $p_{m0} = 0.2$, $p_{M0} = 0.6$, $\lambda_- = -0.1$, $\lambda_+ = 0.1$, and $U_{0B} = 0.6$. Table 1 reports maximum regret computed for various values of ($N_0$, $N_1$, $w_0$) as well as the value of minimax regret, with the optimal weight in parentheses:[5]

---

[4] A Panel developing guidelines for emergency management of ITP has defined a critical bleed to be "a bleed in a critical anatomical site including intracranial, intraspinal, intraocular, retroperitoneal, pericardial, or intramuscular with compartment syndrome; or an ongoing bleed that results in hemodynamic instability or respiratory compromise." See Surotich *et al.* (2021).

[5] The findings in the table were computed as described in Section 3.2.1. At each specified value of ($N_0$, $N_1$, $w_0$), the error probability in a particular state of nature ($p_{s0}$, $p_{s1}$) was approximated by Monte Carlo integration across 20,000 simulated samples. In each pseudo-sample, it was determined whether as-if optimization with the simulated estimate yields an error in treatment choice. The fraction of errors across the 20,000 simulations was used to estimate the error probability and, hence, expected regret. Maximum regret over the state space was approximated by computation of expected regret on a uniform 50 by 50 grid of feasible values for ($p_0$, $p_1$).

Table 1: Maximum Regret with Weighted-Average Estimates

| | $w_0 = 0.5$ | $w_0 = 0.6$ | $w_0 = 0.7$ | $w_0 = 0.8$ | $w_0 = 0.9$ | $w_0 = 1.0$ | MMR (optimal weight) |
|---|---|---|---|---|---|---|---|
| $N_0 = 10, N_1 = 10$ | 0.041 | 0.033 | 0.031 | 0.031 | 0.030 | 0.040 | 0.030 (0.751) |
| $N_0 = 5, \ \ N_1 = 15$ | 0.051 | 0.039 | 0.039 | 0.039 | 0.039 | 0.065 | 0.034 (0.863) |
| $N_0 = 15, N_1 = 5$ | 0.033 | 0.026 | 0.026 | 0.023 | 0.026 | 0.031 | 0.023 (0.752) |
| $N_0 = 20, N_1 = 20$ | 0.033 | 0.026 | 0.023 | 0.022 | 0.021 | 0.026 | 0.021 (0.858) |
| $N_0 = 10, N_1 = 30$ | 0.043 | 0.034 | 0.032 | 0.031 | 0.029 | 0.040 | 0.026 (0.911) |
| $N_0 = 30, N_1 = 10$ | 0.023 | 0.019 | 0.018 | 0.016 | 0.017 | 0.020 | 0.016 (0.800) |

Several features of the findings are noteworthy. First, holding $w_0$ fixed, increasing sample size reduces maximum regret. Doubling both $N_0$ and $N_1$ roughly reduces maximum regret by a factor of $\sqrt{2}$. Second, holding $w_0$ and the total sample size $N_0 + N_1$ fixed, re-allocating sample from $x = 1$ to $x = 0$ always reduces maximum regret. Third, holding $(N_0, N_1)$ fixed, maximum regret is minimized when the weight lies between the polar cases $w_0 = 0.5$ and $w_0 = 1$. The optimal weight is closer to $w_0 = 1$ when sample size is larger.

4.1.3. Estimation by Weighted Averages across Patients with Multiple Covariate Values

Estimation of $p_0$ by a weighted average of outcomes extends easily from the case of a binary covariate to ones where patients have multiple observed covariate values. Let $k = 0, \ldots, K$ index distinct covariate values, with $p_k$ denoting the conditional probability of illness for persons with value $x_k$. For each $k$, let $N_k$ be the number of sampled such patients, which I take to be predetermined, and let $n_k$ be the number observed to be ill.[6]

---

At each specified value of $(N_0, N_1)$, the optimal weight was approximated by computing maximum regret over the uniform grid $w_0 \in [0.50, 0.51, 0.52, \ldots, 0.98, 0.99, 1]$.

[6] To expand on a point made earlier, considering the sample sizes $(N_k, k = 0, \ldots, K)$ to be predetermined simplifies regret analysis relative to a setting where persons are sampled at random from the population at large. In that setting, $(N_k, n_k, k = 0, \ldots, K)$ are jointly random. With the $N_k$ predetermined, the only random variables are $(n_k, k = 0, \ldots, K)$, which are statistically independent of one another. The present analysis applies with sampling at random from the population at large if performance is measured by expected regret conditional on the realized sample sizes rather than by unconditional expected regret.

Let the weights satisfy $0 \leq w_k$ for all k and $\sum_{k=0,\ldots,K} w_k = 1$. Let $w_k N_k > 0$ for at least one value of k. Then a weighted average estimate has the form

$$(35) \quad \varphi_0(n_k, k = 0, \ldots, K) = \frac{\sum\limits_{k=0,\ldots,K} w_k n_k}{\sum\limits_{k=0,\ldots,K} w_k N_k}.$$

Given a specification of the state space, the maximum regret of as-if optimization with estimates of form (35) can be computed numerically and optimal weights determined. For example, one might specify S to satisfy this bounded-variation assumption:

$$(36) \quad p_{sk} + \lambda_{k-} \leq p_{s0} \leq p_{sk} + \lambda_{k+}, \ k = 1, \ldots, K, \ \text{all } s \in S.$$

When the objective is binary treatment choice, analysis of the type sketched here has considerable appeal relative to conventional asymptotic statistical study of kernel estimates. Performance is measured by maximum regret in decision making rather than by maximum mean square error, a concept distant from the decision problem. Exact numerical findings are obtainable for relevant finite sample sizes. In principle, analysis of maximum regret is possible for any specification of the state space. One need not maintain local smoothness assumptions of the types usually imposed in research on kernel estimation of regressions.

### 4.1.4. Bounded Variation Assumptions and the Curse of Dimensionality

I find it interesting to observe that estimation with bounded-variation assumptions is not systematically subject to the *curse of dimensionality*. Traditionally, study of the curse of dimensionality has considered estimation of a regression when the covariates are a finite-dimensional real vector, under a maintained assumption of local smoothness of the regression function. Increasing dimensionality means extending the length of the covariate vector.

The problem is often described by considering a random sample of specified size, each of whose members has an observed J-dimensional real covariate vector x and an observed outcome y. Under usual local smoothness assumptions, the sampling probability with which the covariate value $x_i$ of each observation i lies within a specified Euclidean distance $\varepsilon > 0$ of a covariate value of interest, say $x_0$, is of order $\varepsilon^J$ when $\varepsilon$ is small. Hence, increasing the dimensionality of the covariate space lessens the information that the data yield about the conditional expectation $E(y|x = x_0)$.

In this paper, covariates may lie in a general space, not necessarily a real vector space. To maintain comparability with the traditional setup, let x be a real vector and consider increasing dimensionality. Thus, the extended covariate vector is (x, w), where w is a real vector. Let $(x = x_0, w = w_0)$ be the extended covariate vector of a patient to be treated. Now the objective is to learn $P(y = 1|x = x_0, w = w_0)$ rather than $P(y = 1|x = x_0)$. Whereas the data originally were $[(y_{ki}, x_{ki}), i = 1, \ldots, N_k, k = 0, \ldots, K]$, they now are $[(y_{ki}, x_{ki}, w_{ki}), i = 1, \ldots, N_k, k = 0, \ldots, K]$.

How does dimensional refinement affect estimation with bounded variation assumptions? The answer depends on the applied setting. Whereas inequalities (36) bounded the difference between $P(y = 1|x = x_0)$ and $[P(y = 1|x = x_k), k = 1, \ldots, K]$, a clinician might now seek to credibly bound the difference between $P(y = 1|x = x_0, w = w_0)$ and $[P(y = 1|x = x_k, w = w_k), k = 1, \ldots, K]$. Some of the latter bounds may be tighter than the former ones and others may be looser, depending on the illness and the covariates. Overall, refinement of dimensionality may improve estimation in some applications and weaken it in others.

## 4.2. Ecological Inference

Return to the setting with two covariate values. A bounded-variation assumption directly connects the illness probabilities $p_0$ and $p_1$. A different way to connect $p_0$ and $p_1$ materializes if one has empirical knowledge of the marginal probability of illness in the patient population, say p, and the fractions of the population who have each covariate value, say $r_0$ and $r_1$. For example, suppose that x is a binary measure of obesity and the illness of concern is liver cirrhosis. Public data sources may record the overall rates of obesity and cirrhosis in a population, but not the rate of cirrhosis conditional on obesity status. Then the public data reveal $(p, r_0, r_1)$, but not $(p_0, p_1)$.

Research on the *ecological inference* problem studies the logic of inference on $(p_0, p_1)$ given knowledge of $(p, r_0, r_1)$. The connection among these quantities is shown by the Law of Total Probability

(37)  $p = p_0 r_0 + p_1 r_1$.

Duncan and Davis (1953) observed that (37) implies a computable bound on $p_0$, namely

(38)  $\max[0, (p - r_1)/r_0] \leq p_0 \leq \min(1, p/r_0)$.

The subsequent literature generalizes this finding to settings with general real-valued outcomes. Manski (2018) reviews this work and gives an application to medical decision making.

When evaluating the performance of treatment-choice rules, one may use equation (37) to shrink the state space relative to what it would be in the absence of knowledge of $(p, r_0, r_1)$. Consider any initial state space S, embodying the available restrictions on the true state without knowledge of $(p, r_0, r_1)$. The state space using this knowledge is the non-rectangular set $(s \in S: p = p_{s0} r_0 + p_{s1} r_1)$.

To date, analysis of ecological inference has assumed empirical knowledge only of $(p, r_0, r_1)$, with no sample data on outcomes conditional on covariates. Suppose that one can combine knowledge of $(p, r_0, r_1)$ with observation of outcomes $(n_0, n_1)$ in random samples of sizes $(N_0, N_1)$. Beyond shrinking the state space, one can use (37) when estimating $p_0$. A simple idea emerges by rewriting (37) as $p_0 = (p - p_1 r_1)/r_0$. Thus, if $p_1$ were known, $p_0$ would be known as well. Sample data $(n_0, n_1)$ do not reveal $p_1$ but, when $N_1 > 0$, one may use $n_1/N_1$ to estimate $p_1$. This yields $[p - (n_1/N_1)r_1]/r_0$ as an estimate of $p_0$. This estimate is well-behaved in classical statistical terms, in the sense of being consistent as $N_1 \to \infty$. However, it is inefficient because it does not use the data $n_0$.

To achieve greater precision in estimation, classical statistical thinking suggests minimization of total sample prediction error under square loss, subject to constraint (37). The mean of a probability distribution is the best predictor of a random draw under square loss; hence, $p_0$ and $p_1$ are the best population predictors of illness conditional on $x = 0$ and $x = 1$ respectively. When $(p, r_0, r_1)$ are unknown, this motivates unconstrained least squares estimation, yielding $n_0/N_0$ and $n_1/N_1$ as estimates. When $(p, r_0, r_1)$ are known, it suggests constrained least squares estimation, namely

$$(39) \qquad \min_{(\theta_0, \theta_1) \in [0,1]^2 : \, p = \theta_0 r_0 + \theta_1 r_1} \quad \sum_{i = 1, \ldots, N_0} (y_{0i} - \theta_0)^2 \; + \; \sum_{i = 1, \ldots, N_1} (y_{1i} - \theta_1)^2 .$$

This constrained minimization problem has an explicit solution, which may be interior to or on the boundary of the state space.[7] Solving (37) for $\theta_1$ as a function of $\theta_0$ yields $\theta_1 = \frac{p - \theta_0 r_0}{r_1}$. Inserting this into (39) and solving the first-order condition in $\theta_0$ yields the tentative solution

$$(40) \qquad \theta_0^* = \frac{\sum_{i=1,\ldots,N_0} (Y_{0i}) - \frac{r_0}{r_1} \sum_{i=1,\ldots,N_1} (Y_{1i}) + \frac{p r_0 N_1}{r_1^2}}{\left( N_0 + \frac{N_1 r_0^2}{r_1^2} \right)} .$$

This is the solution if $\max\left\{0, \frac{p - r_1}{r_0}\right\} \leq \theta_0^* \leq \min\left\{1, \frac{p}{r_0}\right\}$. If $\theta_0^* < \max\left\{0, \frac{p - r_1}{r_0}\right\}$, there is a corner solution at $\theta_0 = \max\left\{0, \frac{p - r_1}{r_0}\right\}$. If $\theta_0^* > \min\left\{1, \frac{p}{r_0}\right\}$, there is a corner solution at $\theta_0 = \min\left\{1, \frac{p}{r_0}\right\}$.

Given specified values of $(p_{m0}, p_{M0}, U_{0B}, N_0, N_1, p, r_0, r_1)$, the approximate maximum regret of as-if optimization using the constrained least squares estimate can be computed numerically. To illustrate, let

---

[7] I am grateful to Michael Gmeiner for the derivation, which is obtainable from the author of this paper.

($p_{m0} = 0$, $p_{M0} = 1$, $U_{0B} = \frac{1}{2}$, $p = \frac{1}{2}$, $r_0 = 0.7$, $r_1 = 0.3$) and consider the two sample-size pairs $(N_0, N_1) = (10, 10)$ and $(20, 20)$. Approximating the state space by a grid of 100 values for $p_0$ and maximizing over this grid, the resulting values of maximum regret are 0.011 and 0.008 respectively.

5. Conclusion

This paper carries further my research applying statistical decision theory to treatment choice with sample data, using maximum regret to evaluate the performance of treatment rules. The methodological innovation relative to past work is to study as-if optimization with alternative estimates of illness probabilities, when choosing between surveillance and aggressive treatment. To render the analysis transparent and informative, I studied a relatively simple but decidedly nontrivial formalization of the decision problem. Extending the analysis to more complex and realistic forms of the problem offers much scope for future research.

Beyond the specific analysis performed here, the paper sends a broad message. It is always important to address decision making with care but particularly so in medical settings, where the stakes are often high. Biostatisticians and computer scientists have addressed medical risk assessment in indirect ways, the former applying classical statistical theory and the latter measuring prediction accuracy in test samples. Neither approach is satisfactory. Statistical decision theory provides a coherent, generally applicable methodology.

Appendix: Proofs of Propositions

*Proof of Proposition 1*: The error probability in state s takes the value 0 or 1, with

(A1a)  $Q_s[e(p_{sx}, \varphi_x, U_{xB}) = 1] = 0$  if $\min(1 - p_{sx}, 1 - \varphi_x) \geq U_{xB}$  or  $\max(1 - p_{sx}, 1 - \varphi_x) < U_{xB}$,

(A1b)  $Q_s[e(p_{sx}, \varphi_x, U_{xB}) = 1] = 1$  if  $1 - p_{sx} \geq U_{xB} > 1 - \varphi_x$  or  $1 - p_{sx} < U_{xB} \leq 1 - \varphi_x$.

Expected regret in state s is

(A2a)   $R_{sx}(\varphi_x) = \max [0, (1 - p_{sx}) - U_{xB}]$  if  $U_{xB} > 1 - \varphi_x$,

(A2b)   $R_{sx}(\varphi_x) = \max [0, U_{xB} - (1 - p_{sx})]$  if  $U_{xB} \leq 1 - \varphi_x$.

To compute maximum regret across S, recall the maintained assumption that $p_{mx} \equiv \min_{s \in S} p_{sx} < 1 - U_{xB} < p_{Mx} \equiv \max_{s \in S} p_{sx}$. It follows from (A2a)−(A2b) that

(A3a)   $\max_{s \in S} R_{sx}(\varphi_x) = \max [0, (1 - p_{mx}) - U_{xB}] = (1 - p_{mx}) - U_{xB}$  if  $U_{xB} > 1 - \varphi_x$,

(A3b)   $\max_{s \in S} R_{sx}(\varphi_x) = \max [0, U_{xB} - (1 - p_{Mx})] = U_{xB} - (1 - p_{Mx})$  if  $U_{xB} \leq 1 - \varphi_x$.

These findings yield the estimates that minimize maximum regret.

Q. E. D.

*Proof of Proposition 2*: The error probability in state s with estimate $\varphi_{0x}(\cdot)$ is

(A4)   $Q_o\{e[p_{sx}, \varphi_{0x}(\psi), U_{xB}] = 1\} = Q_0[1 - p_{sx} \geq U_{xB} > 1 - \varphi_{0x}(\psi)$  or  $1 - p_{sx} < U_{xB} \leq 1 - \varphi_{0x}(\psi)]$.

Expected regret is

(A5)   $E_s\{R_{sx}[\varphi_x(\psi)]\} = |(1 - p_{sx}) - U_{xB}|\cdot Q_0\{e[p_{sx}, \varphi_{0x}(\psi), U_{xB}] = 1\}.$

This expression differs from (23) in that the sampling distribution is the known $Q_0$ rather than a state-dependent $Q_s$.

To study maximum regret, partition S into the regions $S_A = (s \in S: 1 - p_{sx} \geq U_{xB})$ and $S_B = (s \in S: 1 - p_{sx} < U_{xB})$. It follows from (A4) and (A5) that

(A6a)   $E_s\{R_{sx}[\varphi_x(\psi)]\} = [(1 - p_{sx}) - U_{xB}]\cdot Q_0[U_{xB} > 1 - \varphi_{0x}(\psi)] = [(1 - p_{sx}) - U_{xB}]\cdot q_{0x}$ for $s \in S_A$,

(A6b)   $E_s\{R_{sx}[\varphi_x(\psi)]\} = [U_{xB} - (1 - p_{sx})]\cdot Q_0[U_{xB} \leq 1 - \varphi_{0x}(\psi)] = [U_{xB} - (1 - p_{sx})]\cdot(1 - q_{0x})$ for $s \in S_B$.

Maximum regret across $s \in S_A$ and $s \in S_B$ are $[(1 - p_{mx}) - U_{xB}]\cdot q_{0x}$ and $[U_{xB} - (1 - p_{Mx})]\cdot(1 - q_{0x})$ respectively. Hence, maximum regret over the entire state space is

(A7)   $\max_{s \in S} E_s\{R_{sx}[\varphi_x(\psi)]\} = \max\{[(1 - p_{mx}) - U_{xB}]\cdot q_{0x}, [U_{xB} - (1 - p_{Mx})]\cdot(1 - q_{0x})\}.$

To minimize maximum regret, consider the right-hand side of (A7) as a function of $q_{0x}$. The expression $[(1 - p_{mx}) - U_{xB}]\cdot q_{0x}$ increases linearly from 0 to $(1 - p_{mx}) - U_{xB}$ as $q_{0x}$ increases from 0 to 1. The expression $[U_{xB} - (1 - p_{Mx})]\cdot(1 - q_{0x})$ decreases linearly from $U_{xB} - (1 - p_{Mx})$ to 0 as $q_{0x}$ increases from 0 to 1. Hence, estimates that minimize maximum regret over $q_{0x} \in [0, 1]$ are those for which $q_{0x}$ solves the equation

(A8)   $[(1 - p_{mx}) - U_x(B)]\cdot q_{0x} = [U_{xB} - (1 - p_{Mx})]\cdot(1 - q_{0x}).$

The unique solution is

$$(A9) \quad q_{0x} = \frac{U_{xB} - (1 - p_{Mx})}{U_{xB} - (1 - p_{Mx}) + [(1 - p_{mx}) - U_{xB}]} = \frac{U_{xB} - (1 - p_{Mx})}{p_{Mx} - p_{mx}} .$$

Inserting this into (A7) gives the minimum achievable value of maximum regret, stated in (28).

<div align="right">Q. E. D.</div>

*Proof of Proposition 3*: The sampling probabilities in state s are $Q_s(0) = 1 - p_{sx}$ and $Q_s(1) = p_{sx}$. Consider the estimate $[\varphi_x(0) = 0, \varphi_x(1) = 1]$. By (25a) – (25b), regret is

$(A10a) \quad E_s\{R_{sx}[\varphi_x(\psi)]\} = [(1 - p_{sx}) - U_{xB}] \cdot p_{sx}$ for $s \in S_A$,

$(A10b) \quad E_s\{R_{sx}[\varphi_x(\psi)]\} = [U_{xB} - (1 - p_{sx})] \cdot (1 - p_{sx})$ for $s \in S_B$.

These expressions are quadratic functions of $p_{sx}$. Examination of first and second-order conditions shows that (A10a) is globally maximized at $(1 - U_{xB})/2$ and (A10b) is globally maximized at $1 - U_{xB}/2$. These maxima lie within $S_A$ and $S_B$ respectively. Hence,

$(A11a) \quad \max_{s \in S_A} E_s\{R_{sx}[\varphi_x(\psi)]\} = [(1 - (1 - U_{xB})/2 - U_{xB}] \cdot (1 - U_{xB})/2 = (1 - U_{xB})^2/4,$

$(A11b) \quad \max_{s \in S_B} E_s\{R_{sx}[\varphi_x(\psi)]\} = (U_{xB} - U_{xB}/2) \cdot (U_{xB}/2) = U_{xB}^2/4.$

Combining (A11a) and (A11b) yields maximum regret over S.

<div align="right">Q. E. D.</div>

*Proof of Proposition 4*: The derivation begins by applying the Law of Iterated Expectations to expected regret in state s. Consider states in $S_A = (s \in S: 1 - p_{sx} \geq U_{xB})$.

$$(A12) \quad E_s[R_{sx}(n_x/N_x)] = E_s[R_{sx}(n_x/N_x) \mid n_x/N_x - p_{sx} \leq \delta] \cdot Q_s(n_x/N_x - p_{sx} \leq \delta)$$
$$+ E_s[R_{sx}(n_x/N_x) \mid n_x/N_x - p_{sx} > \delta] \cdot Q_s(n_x/N_x - p_{sx} > \delta).$$

Observe that $Q_s(n_x/N_x - p_{sx} \leq \delta) \leq 1$ and that $Q_s(n_x/N_x - p_s > \delta) \leq \exp(-2N_x\delta^2)$ by the Hoeffding large-deviation inequality. Also observe that $R_{sx}(n_x/N_x) \leq (1 - p_{sx}) - U_{xB}$. Combining these results with (A12) gives this upper bound on expected regret:

$$(A13) \quad E_s[R_{sx}(n_x/N_x)] \leq E_s[R_{sx}(n_x/N_x) \mid n_x/N_x - p_{sx} \leq \delta] + [(1 - p_{sx}) - U_{xB}] \cdot \exp(-2N_x\delta^2).$$

It remains to consider $E_s[R_{sx}(n_x/N_x) \mid n_x/N_x - p_{sx} \leq \delta]$. When $(1 - p_{sx}) - U_{xB} > \delta$, the condition $n_x/N_x - p_{sx} \leq \delta$ implies that $(1 - n_x/N_x) - U_{xB} > 0$. Hence, $e(p_{sx}, n_x/N_x, U_{xB}) = 0$ and $R_{sx}(n_x/N_x) = 0$. When $(1 - p_{sx}) - U_{xB} \leq \delta$, then $R_{sx}(n_x/N_x) \leq \delta$. Hence, $E_s[R_{sx}(n_x/N_x) \mid n_x/N_x - p_{sx} \leq \delta] \leq \delta$. Combining this inequality with (A13) yields

$$(A14) \quad E_s[R_{sx}(n_x/N_x)] \leq \delta + [(1 - p_{sx}) - U_{xB}] \cdot \exp(-2N_x\delta^2).$$

Finally, maximizing expected regret over $S_A$ yields

$$(A15) \quad \max_{s \in S_A} E_s[R_{sx}(n_x/N_x)] \leq \delta + [(1 - p_{mx}) - U_{xB}] \cdot \exp(-2N_x\delta^2).$$

Now consider states in $S_B = (s \in S: 1 - p_{sx} < U_{xB})$. An analogous derivation yields

(A16)      $\max_{s \in S_B}$  $E_s[R_{sx}(n_x/N_x) \leq \delta + [U_{xB} - (1 - p_{Mx})] \cdot \exp(-2N_x\delta^2)$.

Combining (A15) and (A16) yields result (30).

Q. E. D.

*Proof of Proposition 5*: The sampling probabilities in state s are $Q_s(n_1 = 0) = 1 - p_{s1}$ and $Q_s(n_1 = 1) = p_{s1}$.
Regret is

(A17a)   $E_s[R_{s0}(n_1)] = [(1 - p_{s0}) - U_{0B}] \cdot p_{s1}$  for $s \in S_A$,

(A17b)   $E_s[R_{s0}(n_1)] = [U_{0B} - (1 - p_{s0})] \cdot (1 - p_{s1})$ for $s \in S_B$.

Consider $s \in S_{0A}$; thus, $p_{s0} \leq 1 - U_{0B}$. By assumption, $\lambda \leq \min(U_{0B}, 1 - U_{0B})$. Hence, $p_{s0} + \lambda \leq 1$. Holding $p_{s0}$ fixed, maximum expected regret in (A17a) over $p_{s1}$, subject to (31), occurs when $p_{s1} = p_{s0} + \lambda$. This yields maximum regret $(1 - p_{s0} - U_{0B}) \cdot (p_{s0} + \lambda)$, a quadratic function of $p_{s0}$ alone. Examination of first and second-order conditions shows that (A17a) is globally maximized at $(1 - U_{0B} - \lambda)/2$. This maximum lies within $S_{0A}$. The value of the maximum across $S_A$ is $[(1 - U_{0B}) + \lambda]^2/4$.

Consider $s \in S_{0B}$; thus, $p_{s0} > 1 - U_{0B}$. By assumption, $\lambda \leq \min(U_{0B}, 1 - U_{0B})$. Hence, $p_{s0} - \lambda > 0$. Holding $p_{s0}$ fixed, maximum expected regret in (A17b) over $p_{s1}$, subject to (31), occurs when $p_{s1} = p_{s0} - \lambda$. This yields maximum regret $[U_{0B} - (1 - p_{s0})] \cdot (1 - p_{s0} + \lambda)$, a quadratic function of $p_{s0}$ alone. Examination of first and second-order conditions shows that (A17b) is globally maximized at $1 - U_{0B}/2 + \lambda/2$. This maximum lies within $S_{0B}$. The value of the maximum across $S_B$ is $(U_{0B} + \lambda)^2/4$.

Combining the findings for $S_{0A}$ and $S_{0B}$ yields maximum regret over S.

Q. E. D.

*Proof of Proposition 6*: In state s, the estimate $(n_0 + n_1)/(N_0 + N_1)$ has mean $\alpha_0 p_{s0} + \alpha_1 p_{s1}$, where $\alpha_0 \equiv N_0/(N_0 + N_1)$ and $\alpha_1 \equiv N_1/(N_0 + N_1)$. Hence, by (31),

(A18)  $|p_{s0} - (\alpha_0 p_{s0} + \alpha_1 p_{s1})| = \alpha_1 |p_{s0} - p_{s1}| \le \alpha_1 \lambda.$

The large-deviations inequality of Hoeffding (1963) shows that, for all $\delta \in (0, 1)$ and $s \in S$,

(A19a)  $Q_s[(n_0 + n_1)/(N_0 + N_1) - (\alpha_0 p_{s0} + \alpha_1 p_{s1}) > \delta] \le \exp[-2(N_0 + N_1)\delta^2],$

(A19b)  $Q_s[(\alpha_0 p_{s0} + \alpha_1 p_{s1}) - (n_0 + n_1)/(N_0 + N_1) > \delta] \le \exp[-2(N_0 + N_1)\delta^2].$

Combining this with (A18) yields

(A20a)  $Q_s[(n_0 + n_1)/(N_0 + N_1) - p_{s0} > \delta + \alpha_1\lambda] \le \exp[-2(N_0 + N_1)\delta^2],$

(A20b)  $Q_s[p_{s0} - (n_0 + n_1)/(N_0 + N_1) > \delta + \alpha_1\lambda] \le \exp[-2(N_0 + N_1)\delta^2].$

The rest of the proof is similar to the proof to Proposition 4, with $\delta + \alpha_1\lambda$ replacing $\delta$ when the Law of Iterated Expectations is used to decomposed expected regret. Consider $s \in S_{0A} = (s \in S: 1 - p_{s0} \ge U_{0B})$.

(A21)  $E_s\{R_{s0}[(n_0 + n_1)/(N_0 + N_1)]\} =$

$E_s\{R_{s0}[(n_0 + n_1)/(N_0 + N_1)] \mid (n_0 + n_1)/(N_0 + N_1) - p_{s0} \le \delta + \alpha_1\lambda\} \cdot Q_s[(n_0 + n_1)/(N_0 + N_1) - p_{s0} \le \delta + \alpha_1\lambda]$

$+ E_s\{R_{s0}[(n_0 + n_1)/(N_0 + N_1)] \mid (n_0 + n_1)/(N_0 + N_1) - p_{s0} > \delta + \alpha_1\lambda\} \cdot Q_s[(n_0 + n_1)/(N_0 + N_1) - p_{s0} > \delta + \alpha_1\lambda].$

$Q_s[(n_0 + n_1)/(N_0 + N_1) - p_{s0} \le \delta + \alpha_1\lambda] \le 1.$ $R_{s0}[(n_0 + n_1)/(N_0 + N_1)] \le (1 - p_{s0}) - U_{0B}$ for all values of $(n_0 + n_1)/(N_0 + N_1)$. Combining these results with (A20a) gives this upper bound on expected regret:

(A22)   $E_s\{R_{s0}[(n_0 + n_1)/(N_0 + N_1)]\} \leq E_s\{R_{s0}[(n_0 + n_1)/(N_0 + N_1)] \mid (n_0 + n_1)/(N_0 + N_1) - p_{s0} \leq \delta + \alpha_1\lambda\}$

$+ [(1 - p_{s0}) - U_{0B}]\cdot exp[-2(N_0 + N_1)\delta^2]$.

Now consider $E_s\{R_{s0}[(n_0 + n_1)/(N_0 + N_1)] \mid (n_0 + n_1)/(N_0 + N_1) - p_{s0} \leq \delta + \alpha_1\lambda\}$. When $(1 - p_{s0}) - U_{0B}$ $> \delta + \alpha_1\lambda$, the condition $(n_0 + n_1)/(N_0 + N_1) - p_{s0} \leq \delta + \alpha_1\lambda$ implies that $[1 - (n_0 + n_1)/(N_0 + N_1)] - U_{0B} > 0$. Hence, $e[p_{s0}, (n_0 + n_1)/(N_0 + N_1), U_{0B}] = 0$ and $R_{s0}[(n_0 + n_1)/(N_0 + N_1)] = 0$. When $(1 - p_{s0}) - U_{0B} \leq \delta + \alpha_1\lambda$, then $R_{s0}[(n_0 + n_1)/(N_0 + N_1)] \leq \delta + \alpha_1\lambda$. Hence, $E_s\{R_{s0}[(n_0 + n_1)/(N_0 + N_1)] \mid (n_0 + n_1)/(N_0 + N_1) - p_{s0} \leq \delta + \alpha_1\lambda\} \leq \delta + \alpha_1\lambda$. Combining this inequality with (A22) yields

(A23)   $E_s\{R_{s0}[(n_0 + n_1)/(N_0 + N_1)]\} \leq (\delta + \alpha_1\lambda) + [(1 - p_{s0}) - U_{0B}]\cdot exp[-2(N_0 + N_1)\delta^2]$.

Maximizing expected regret over $S_A$ yields

(A24)       $\max_{s \in S_A} E_s\{R_{s0}[(n_0 + n_1)/(N_0 + N_1)]\} \leq (\delta + \alpha_1\lambda) + [(1 - p_{m0}) - U_{0B}]\cdot exp[-2(N_0 + N_1)\delta^2]$.

For states in $S_B = (s \in S: 1 - p_{s0} < U_{0B})$, an analogous derivation yields

(A25)       $\max_{s \in S_B} E_s\{R_{s0}[(n_0 + n_1)/(N_0 + N_1)]\} \leq (\delta + \alpha_1\lambda) + [U_{0B} - (1 - p_{M0})]\cdot exp[-2(N_0 + N_1)\delta^2]$.

Combining (A24) and (A25) yields (33).

Q. E. D.